\documentclass[prb,preprint,preprintnumbers,amsmath,amssymb]{revtex4}
\usepackage{mathrsfs}
\usepackage{graphicx}% Include figure files

\begin{document}
%title
\title{Kondo Effect in a Spin-3/2 Fermi Gas}

\author{Bei Xu, Shoufa Sun, and Qiang Gu}\email{qgu@ustb.edu.cn}
\affiliation{Department of Physics, University of Science and
Technology Beijing, Beijing 100083, China}

\date{\today}
	
\begin{abstract}
	We investigate the Kondo effect of a spin-3/2 Fermi gas and give a detailed calculation of the impurity resistance and ground state energy based on the s-d exchange model. It is found that the impurity resistance increases logarithmically with the decrease of temperature in the case of antiferromagnetic coupling similar to the spin-1/2 system but has a larger resistance minimum value due to the increase of spin scattering channels. In the case of antiferromagnetic interaction, the ground state is still the Kondo singlet state while the septuplet state has the lowest energy for ferromagnetic coupling. And with the same antiferromagnetic s-d coupling parameter, the energy of the Kondo singlet state is lower than spin-1/2, which indicates that the larger spin, the easier it is to enter the Kondo-screened phase. This provides some theoretical support for the realization of the Kondo effect with ultra-cold atoms.
\end{abstract}
	
%\keywords{large-spin, Kondo effect, s-d model, Kondo singlet state}

\maketitle	
	
\section{Introduction}
	
The Kondo effect that includes a series of low-temperature anomalies such as minimum resistance, an anomaly of susceptibility, and specific heat capacity reveals the correlation between a magnetic impurity and itinerant electrons \cite{Hewson1997,Kondo1964}. And the cause of Kondo effect in dilute magnetic alloys can be perfectly described by the s-d exchange model \cite{Kondo1964} or the Anderson single impurity model \cite{Anderson1961}. And the Kondo problem can be reduced to a one-dimensional Ising model with inverse-square long-range interaction in some limit case \cite{Anderson1971}. Besides that, below the Kondo temperature $T_k$, the local spin forms a spin-singlet state with conduction electrons via antiferromagnetic exchange interaction, which is viewed as a fully Kondo screening state \cite{Yosida1966,Wilson1975,Andrei1980}.

In addition to dilute magnetic alloys, many fascinating Kondo phenomena have been observed in some new structures or materials such as quantum dots \cite{GG1998,Cron1998,Choi2004,Keller2014,Weymann2018}, carbon nanotube \cite{Nygard2000,Pablo2005,Thiago2006,Thiago2020}, superconductors \cite{Balatsky2006,Franke2011,Michael2015,Benjamin2018,Liu2019}, and graphene \cite{Fritz2013,Lo2014,David2017,Diniz2018,Li2019}. For instance, quantum dots provide a good artificial platform to adjust the coupling parameters of the Kondo effect by controlling the depth of the potential well. The existence of Cooper pairs in superconductors creates an energy gap on the Fermi surface, resulting in the instability of the Kondo singlet state of magnetic impurity. And the competition between superconducting and Kondo correlations leads to the emergence of some new phase states. Especially in the topological superconductors, the intrinsic and extrinsic case shows distinct Kondo ground states driven by different pairing mechanisms \cite{Wang2019}. Moreover, the graphene with exotic Dirac-like electron excitations produces rich and prominent pseudogap Kondo problems.

Furthermore, with the realization of degenerate Fermi gas \cite{DSJin1999,Fukuhara2007}, ultracold Fermi atoms are also proposed to offer implementations for the Kondo or Anderson impurity model \cite{Foss2010,Bloch2012,Bauer2013,Bauer2015,Kuzmenko2015,Nakagawa2015}. And a rich variety of exotic Kondo physics can be simulated with large spin Fermi gases by precise control of spin-exchange interactions \cite{Scazza2014,Kuzmenko2016Yb,Zhai2016,Zhai2017,Nagy2018,Zhai2019,Mi2019,Zhai2021}. For example, the Yb atoms have been successfully used to realize the two-orbital Kondo model and there emerges non-Fermi liquid behavior at low tempreture \cite{Riegger2018,Kuzmenko2018}. For isotopes with $S \geq 3/2$, it can allow for the realization of models with SU(N) symmetric form \cite{Miguel2014}. The theoretical proposal and experimental design of the SU (N) symmetric Kondo model are roughly divided into two schemes: multichannel \cite{Lal2010,Nishida2013,Nishida2016} and single-channel with large spin \cite{Gors2010,Nakagawa2015,Zhai2016,Riegger2018}. Particularly the complicated spin-exchange collision between different hyperfine states plays an important role in the large spin Fermi gas \cite{Scazza2014,Koki2019,Zhai2020}, so we pay attention to the detailed dependence of the impurity resistance and Kondo ground state on the spin-exchange collisions in spin-3/2 Fermi gas.

In the paper, we will consider the case of fully screening, utilize the perturbation theory to analyze the Kondo scattering mechanism of spin-3/2 Fermi gas based on the extended s-d exchange model. We calculate the scattering probability and the ground state for the antiferromagnetic and ferromagnetic cases separately. We find that the resistance has logarithmic singularity with decreasing temperature, which is similar to the spin-1/2 system except for its larger coefficient magnitude. For the ground state of the system, septuplet state is the  lowerest energy state in the case of ferromagnetic coupling, while the Kondo singlet state is the most stable under antiferromagnetic coupling analogous to the SU(N) Kondo problem \cite{Koki2019}. The remainder of this paper is organized as follows. In Sec. 2, we present the Kondo Hamiltonian of the large spin Fermi gas with S=3/2 and calculate the scattering probability of the itinerant fermionic atoms to the second Born approximation. In Sec. 3, we give a derivation of impurity resistance and Kondo ground state energy of the system. At last, a brief conclusion and an outlook on promising future work are made.
			
\section{Model and Method} \label{2}
\subsection{The Kondo Hamiltonian of spin-3/2 Fermi gases}

Analogous to the electronic system with spin S = 1/2, the Kondo Hamiltonian of the large-spin fermions with spin S = 3/2 can be expressed as follows
\begin{eqnarray}	
	\hat{H}=\sum_{k\mu}\varepsilon_{k}c^{\dagger}_{k\mu}c_{k\mu} -\frac{J}{N} \sum_{k k^{'}\mu\mu^{'}}\hat{S}\cdot\hat{\sigma}_{\mu\mu^{'}}c^{\dagger}_{k\mu}c_{k^{'}\mu^{'}},
\end{eqnarray}	
where $c^{\dagger}_{k\mu}(c_{k\mu})$ creates (annihilates) a fermi atom with momentum $k$ and spin $\mu=\pm3/2,\pm1/2$, $\varepsilon_{k}$ is the kinetic energy of the itinerant atom and it is independent of spin. The second term describes the s-d exchange interaction between the itinerant fermionic atoms and a localized impurity atom with spin S=3/2. $\hat{S}$ represents the spin operator of the localized impurity atom. $\hat{\sigma}$ describes the spin of the itinerant fermionic atoms. And
\begin{eqnarray}
	\hat{S}\cdot\hat{\sigma}_{\mu\mu^{'}}=\hat{S}^{z}\hat{\sigma}^{z}+\frac{1}{2}(\hat{S}^{+}\hat{\sigma}^{-}+\hat{S}^{-}\hat{\sigma}^{+}),
\end{eqnarray}
$\hat{S}^{\pm}$ and $\hat{\sigma}^{\pm}$ respectively represent the spin lift (down) operators of the impurity atom and the itinerant atoms.
The matrix representation of Pauli spin operator for the spin-3/2 particle is
\begin{eqnarray}\label{pauli}
	&&	\sigma^z=\left(\begin{array}{cccc}
		3&0&0&0\\
		0&1&0&0\\
		0&0&-1&0\\
		0&0&0&-3	
	\end{array}
	\right),\nonumber \\
	&&	\sigma^{\dagger}=\left(\begin{array}{cccc}
		0&\sqrt{3}&0&0\\
		0&0&2&0\\
		0&0&0&\sqrt{3}\\
		0&0&0&0	
	\end{array}
	\right), \nonumber\\
	&&	\sigma^-=\left(\begin{array}{cccc}
		0&0&0&0\\
		\sqrt{3}&0&0&0\\
		0&2&0&0\\
		0&0&\sqrt{3}&0	
	\end{array}
	\right).
\end{eqnarray}
And the spin operator of impurity atom satisfies the commutation relation $[\hat{S}^{+},\hat{S}^{-}]=2\hat{S}^{z};[\hat{S}^{z},\hat{S}^{+}]=\hat{S}^{+};[\hat{S}^{-},\hat{S}^{z}]=\hat{S}^{-}$.
Then we consider the case that the coupling strength $|J|$ is much smaller than the Fermi energy of the itinerant Fermi atom( $|J|\ll E_F$). We use perturbation theory to calculate the scattering probability of the itinerant atom from the initial state $(k,\mu)$ to the final state $(k^{'},\mu^{'})$ and simultaneously the impurity spin state from $\alpha$ to $\alpha^{'}$.

\subsection{Scattering probability}

The first-order scattering amplitude can be written as
\begin{eqnarray}
	T^{(1)}_{\mu^{'}\alpha^{'},\mu\alpha}=-\left(\frac{J}{N}\right)\hat{\sigma}_{\mu^{'}\mu}\cdot\hat{S}_{\alpha^{'}\alpha},
\end{eqnarray}
and in the representation of itinerant atoms, the corresponding transition matrix is expressed as
\begin{eqnarray}
	T^{(1)}=-\left(\frac{J}{N}\right)\left(\begin{array}{cccc}
		3\hat{S}^{z}&\sqrt{3}\hat{S}^{+}&0&0\\
		\sqrt{3}\hat{S}^{-}&\hat{S}^{z}&2\hat{S}^{+}&0\\
		0&2\hat{S}^{-}&-\hat{S}^{z}&\sqrt{3}\hat{S}^{+}\\
		0&0&\sqrt{3}\hat{S}^{-}&-3\hat{S}^{z}	
	\end{array}
	\right).
\end{eqnarray}
Then under the Born approximation, the scattering probability is proportional to
\begin{eqnarray}
	|T^{(1)}|^2=20\left(\frac{J}{N}\right)^2S(S+1),
\end{eqnarray}
which is ten times that of spin-1/2 systems and independent of temperature.
Complications arise in the second order. For the s-d model, the change of z component of the impurity spin angular momentum is restricted to 0, $\pm$1. Except for some forbidden spin-flip scattering channels, there exist fourteen groups of second-order scattering processes in spin-3/2 Fermi gas. The itinerant atoms scatter with the local impurity through an intermediate state $(k''\mu'')$. And the occupation of the intermediate state satisfies the Fermi distribution function,
\begin{eqnarray}
	f_{k''}=[exp(\frac{\epsilon_{k''}}{k_{B}T})+1]^{-1}.
\end{eqnarray}	
Take the atom with the initial state $(k,3/2)$ is scattered to the final state $(k',3/2)$ as an example and the intermediate state is $(k''\mu'')$. The scattering process $<k \frac{3}{2}|T_{k k^{'}}|k^{'} \frac{3}{2}>$ can be divided into four groups as shown below:	
(1) The atom in $(k,3/2)$ is first scattered to the unoccupied state $(k'',3/2)$ and then scattered to $(k',3/2)$ as shown in Fig.1 (a). The corresponding matrix element is
\begin{eqnarray}
	9\left(\frac{J}{N}\right)^{2}\sum_{k''}\frac{(\hat{S}^{z})^2}{\epsilon-\epsilon_{k''}}(1-f_{k''\frac{3}{2}}).
\end{eqnarray}
(2) A atom in an occupied state $(k'',3/2)$ is scattered into the state $(k',3/2)$, and the remaining hole $(k'',3/2)$ is annihilated by the initial $(k,3/2)$ shown as Fig.1 (b). The corresponding matrix element is
\begin{eqnarray}
	9\left(\frac{J}{N}\right)^{2}\sum_{k''}\frac{(\hat{S}^{z})^2}{\epsilon-\epsilon_{k''}}f_{k''\frac{3}{2}}.
\end{eqnarray}	
% Figure
\begin{figure}
	\centering
	\includegraphics[width=0.35\textwidth]{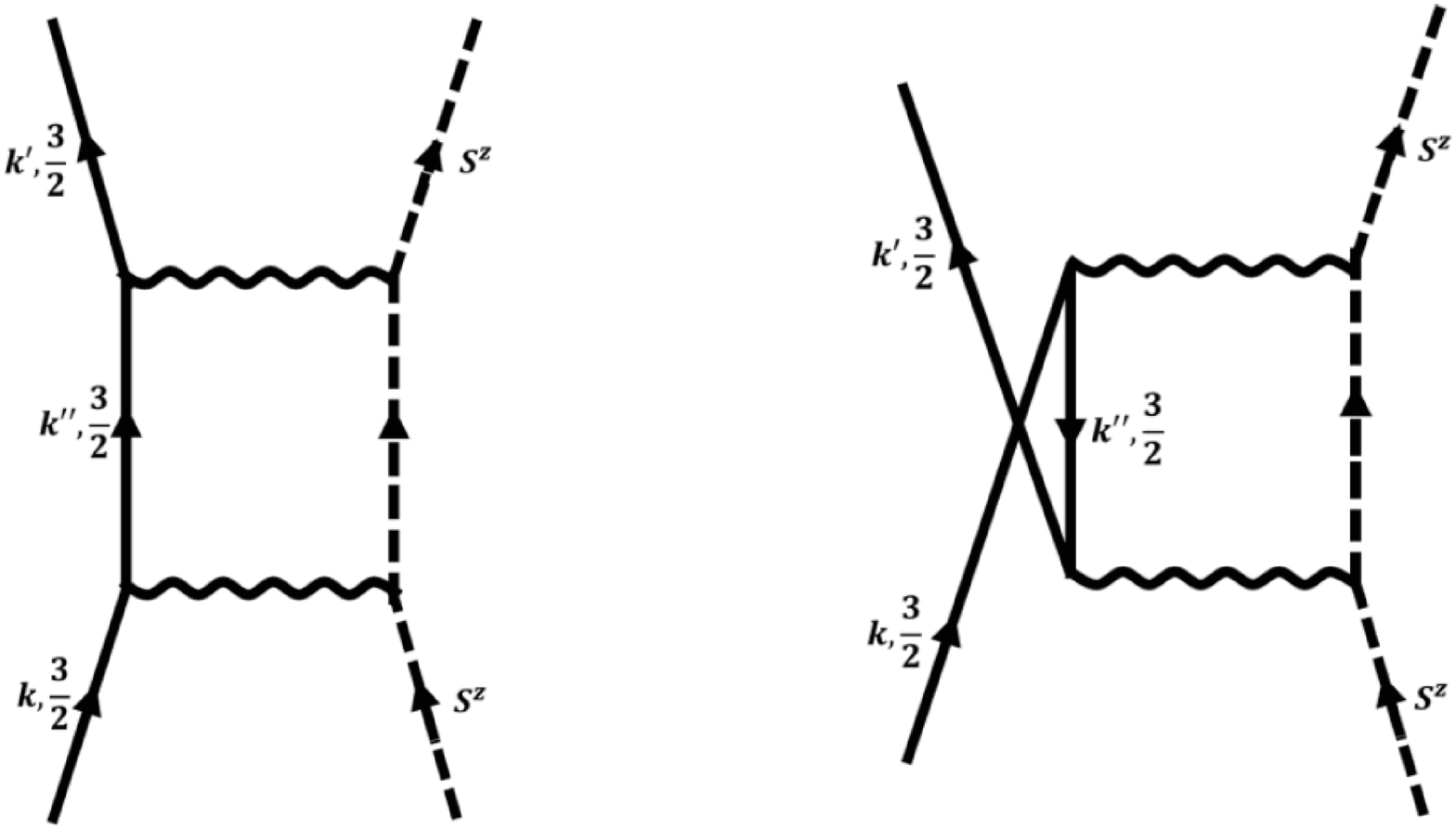}
	\caption{Second-order scattering from $(k,\frac{3}{2})$ to $(k',\frac{3}{2})$ with the intermediate state $(k'',\frac{3}{2})$. (a) unoccupied intermediate state. (b) occupied intermediate state. Solid lines show particle lines, dashed lines show interaction lines, and arrows show scattering directions.}
\end{figure}	
(3) Spin changes in the intermediate state. The atom with $(k,3/2)$ is scattered to the state $(k'',1/2)$ and then to the state $(k',3/2)$ shown as in Fig.2 (a). This scattering proscess is absent in the spin-1/2 electron system. The corresponding matrix element is
\begin{eqnarray}
	3\left(\frac{J}{N}\right)^{2}\sum_{k''}\frac{\hat{S}^{-}\hat{S}^{+}}{\epsilon-\epsilon_{k''}}(1-f_{k''\frac{1}{2}}).
\end{eqnarray}
(4) The scattering process when the intermediate state $(k'',1/2)$ is occupied is shown as Fig.2 (b). The corresponding matrix element is
\begin{eqnarray}\label{T3/2}
	3\left(\frac{J}{N}\right)^{2}\sum_{k''}\frac{\hat{S}^{+}\hat{S}^{-}}{\epsilon-\epsilon_{k''}}f_{k''\frac{1}{2}}.
\end{eqnarray}
\begin{figure}
	\centering
	\includegraphics[width=0.35\textwidth]{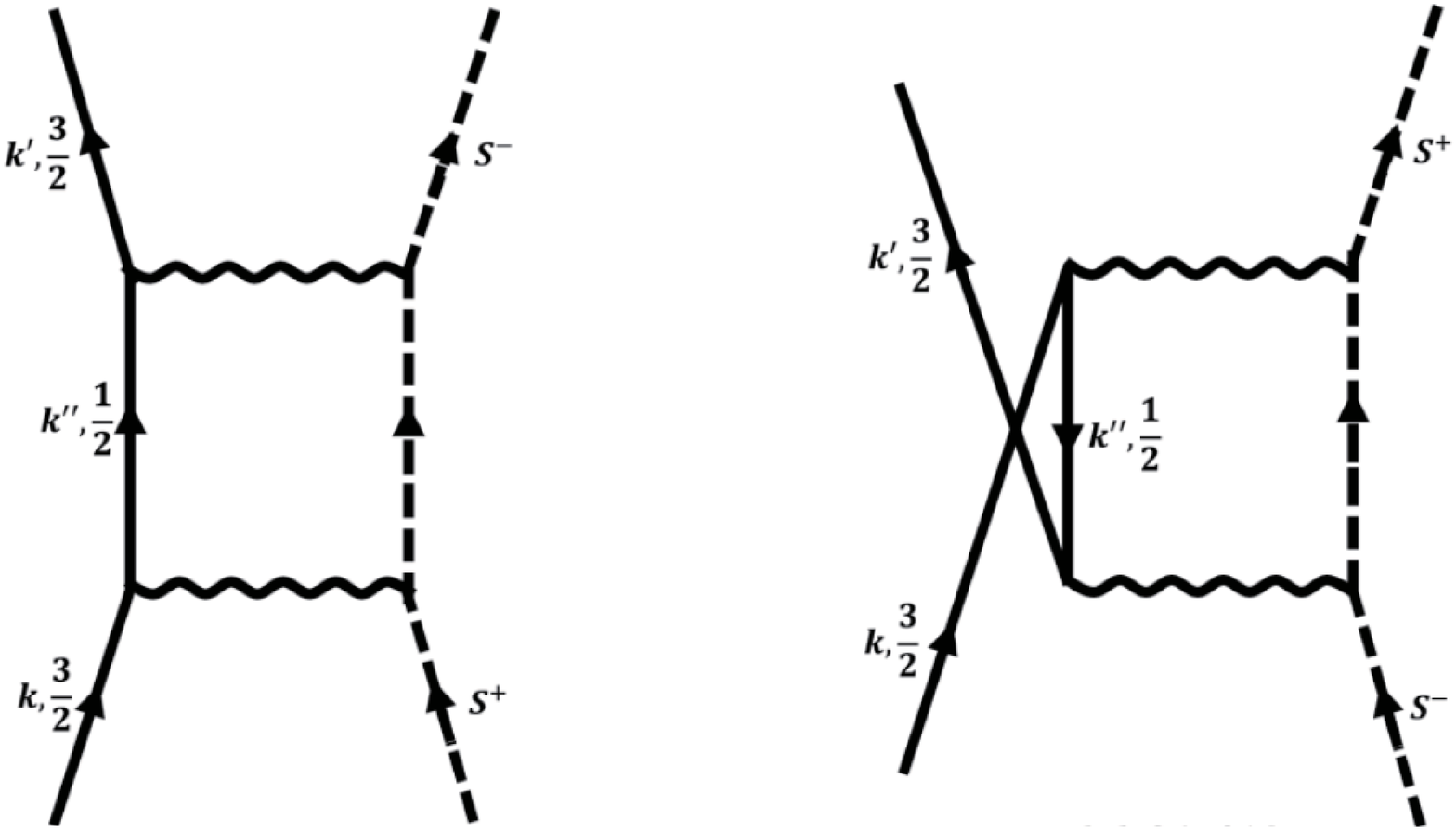}
	\caption{Second-order scattering from $(k,\frac{3}{2})$ to $(k',\frac{3}{2})$ with the intermediate state $(k'',\frac{1}{2})$. (a) unoccupied intermediate state. (b) occupied intermediate state. Solid lines show particle lines, dashed lines show interaction lines, and arrows show scattering directions.
	}
\end{figure}	
So the second-order scattering amplitude of atom from the initial state $(k,3/2)$ to the final state $(k^{'},3/2)$ can be written as
\begin{eqnarray}
	T_{\frac{3}{2}\alpha^{'},\frac{3}{2}\alpha}^{(2)}(k',k)=3\left(\frac{J}{N}\right)^{2}\sum_{k''}( \frac{3(\hat{S}^{z})^2+\hat{S}^{-}\hat{S}^{+}}{\epsilon-\epsilon_{k''}}
	+[\hat{S}^{+},\hat{S}^{-}]\frac{f_{k''\frac{1}{2}}}{\epsilon-\epsilon_{k''}}).
\end{eqnarray}
The two terms in Equation {\ref{T3/2}} have quite different behavior. The first term has no Fermi factor and is irrelevant to temperature. The remaining term contributes to the temperature-dependent magnetic impurity resistance. Usually, it is assumed that the energy of both the incident state and the outgoing state of itinerant particles is equal to 0 ($\epsilon_{k}=\epsilon_{k'}=\epsilon =0$). Since the kinetic energy of the itinerant atom is spin-independent, the intermediate state is degenerate with the same distribution function. Then, the scattering amplitude is evaluated by using the commutation relation of the spin operator and changing the summation over $k''$ to integrations over $\epsilon_{k''}$,
\begin{eqnarray}
	T^{(2)}_{\frac{3}{2} \alpha^{'},\frac{3}{2} \alpha}\approx 3\left(\frac{J}{N}\right)\hat{S}^{z}(2J\rho_{F} ln\frac{k_{B}T}{D}),	
\end{eqnarray}
where $\rho_F$ and D express the average state density per particle on the Fermi surface and the bandwidth.
Other temperature-dependent second-order scattering terms can be obtained similarly. The scattering matrix elements corresponding to different initial and final states are expressed as
\begin{eqnarray}
	&&T_{\frac{1}{2}\alpha^{'},\frac{3}{2}\alpha}\approx \sqrt{3}\left(\frac{J}{N}\right)\hat{S}^{+}(2J\rho_{F} ln\frac{k_{B}T}{D}); \nonumber\\
	&&T_{-\frac{1}{2}\alpha^{'},\frac{3}{2}\alpha}=0; \nonumber\\
	&&T_{\frac{3}{2}\alpha^{'},\frac{1}{2}\alpha}\approx \sqrt{3}\left(\frac{J}{N}\right)\hat{S}^{-}(2J\rho_{F} ln\frac{k_{B}T}{D}); \nonumber\\
	&&T_{\frac{1}{2}\alpha^{'},\frac{1}{2}\alpha}\approx \left(\frac{J}{N}\right)\hat{S}^{z}(2J\rho_{F} ln\frac{k_{B}T}{D}); \nonumber\\
	&&T_{-\frac{1}{2}\alpha^{'},\frac{3}{2}\alpha}\approx 2\left(\frac{J}{N}\right)\hat{S}^{+}(2J\rho_{F} ln\frac{k_{B}T}{D}); \nonumber\\
	&&T_{-\frac{3}{2}\alpha^{'},\frac{1}{2}\alpha}=0; \nonumber\\
	&&T_{\frac{3}{2}\alpha^{'},-\frac{1}{2}\alpha}=0; \nonumber\\
	&&T_{\frac{1}{2}\alpha^{'},-\frac{1}{2}\alpha}\approx 2\left(\frac{J}{N}\right)\hat{S}^{-}(2J\rho_{F} ln\frac{k_{B}T}{D}); \nonumber\\
	&&T_{-\frac{1}{2}\alpha^{'},-\frac{1}{2}\alpha}\approx -\left(\frac{J}{N}\right)\hat{S}^{z}(2J\rho_{F} ln\frac{k_{B}T}{D}); \nonumber\\
	&&T_{-\frac{3}{2}\alpha^{'},-\frac{1}{2}\alpha}\approx \sqrt{3}\left(\frac{J}{N}\right)\hat{S}^{+}(2J\rho_{F} ln\frac{k_{B}T}{D}); \nonumber\\
	&&T_{\frac{1}{2}\alpha^{'},-\frac{3}{2}\alpha}=0; \nonumber\\
	&&T_{-\frac{1}{2}\alpha^{'},-\frac{3}{2}\alpha}\approx \sqrt{3}\left(\frac{J}{N}\right)\hat{S}^{-}(2J\rho_{F} ln\frac{k_{B}T}{D}); \nonumber\\
	&&T_{-\frac{3}{2}\alpha^{'},-\frac{3}{2}\alpha}\approx -3\left(\frac{J}{N}\right)\hat{S}^{z}(2J\rho_{F} ln\frac{k_{B}T}{D}).
\end{eqnarray}

Finally, by adding all these terms together, we can get the total scattering probability of the system,
\begin{eqnarray}
	&&	|T^{(1)}+T^{(2)}|^2=20S(S+1)\left(\frac{J}{N}\right)^{2}(1+4J\rho_{F}ln\frac{k_{B}T}{D}). \nonumber\\
	&&		
\end{eqnarray}

\section{Results and discussions}
\subsection{Impurity resistivity}

The total impurity resistivity should be proportional to the probability that the itinerant particles are scattered by the impurity on the Fermi surface ($\epsilon =0$). And the scattering probability in the above produces a term in the resistivity that goes as $ln T$.
The inverse of relaxation time $1/\tau_k$ is written as	
\begin{eqnarray} \label{tk}
	&&\frac{1}{\tau_k}= 2\pi n_{imp}\sum_{\mu\mu^{'}}\int\frac{d \stackrel{\rightarrow}{k^{'}}}{(2\pi)^3} \nonumber\\
	&&\left[\delta(\epsilon_{k}-\epsilon_{k^{'}}) |T_{\mu^{'} k^{'},\mu k}^{(1)}+T_{\mu^{'} k^{'},\mu k}^{(2)}|^2 (1-\cos\theta)\right] \nonumber\\
	&&= 20S(S+1)(1+4J\rho_{F}ln\frac{k_{B}T}{D})\frac{3n\cdot n_{imp}\cdot J^2}{2e^2\hbar\epsilon_{F}},
\end{eqnarray}
where $n_{imp}$ is the density of magnetic impurity and $n$ is the density of itinerant atoms.	
And the standard expression for the resistance is \cite{Hewson1997}
\begin{eqnarray}\label{R}
	R= \frac{m^{*}}{ne^2\tau_{k}},
\end{eqnarray}
where $m^{*}$ is the effective mass of the itinerant atom.
Substitute Eq. {\ref{tk} into Eq. {\ref{R}}, then
	\begin{eqnarray}
		&&R=\frac{3m^{*}\cdot n_{imp}\cdot J^2}{2e^4\hbar\epsilon_{F}}20S(S+1)(1+4J\rho_{F}ln\frac{k_{B}T}{D}).\nonumber\\
		&&
	\end{eqnarray}
	The factor $ln (k_{B}T/D)$ is negative at low temperatures. And with antiferromagnetic coupling $J<0$, the resistance value increases as the Kondo term becomes larger. It predicts a minimum in resistivity. In addition, the spin-flip scattering process between the itinerant atoms and the local magnetic moment is more frequent due to the multiple spin components. As a result, the impurity resistance of spin-3/2 Fermi gas is ten times that of the spin-1/2 system as the temperature $T\to0$.	
	
\subsection{Ground state energy}
	
	In this section, we will study the ground state of the spin-3/2 Fermi system consisting of the itinerant fermionic atoms and a localized impurity atom under ferromagnetic and antiferromagnetic coupling separately based on the extended s-d exchange model. For the spin-1/2 electron system, the impurity electron forms a collective spin state with the itinerant electrons at low temperature, which can be viewed as a Kondo screening effect. Next, we will give a derivation of the bound state energies for the case S=3/2 by using the method introduced by Yosida in 1966 \cite{Yosida1966} and consider the influence of the spin-mixing interactions between large spin Fermi atoms on the fully screened ground states.	
	At T=0K, the collective spin states can be formed by adding an impurity atom above the Fermi sea. The simplified Hamiltonian of the system is expressed as
	\begin{eqnarray}	
		&&\hat{H}=\sum_{k\mu}\varepsilon_{k}c^{\dagger}_{k\mu}c_{k\mu} -\frac{J}{N} \sum_{k k^{'}>k_F\mu\mu^{'}}\hat{S}\cdot\hat{\sigma}_{\mu\mu^{'}}c^{\dagger}_{k\mu}c_{k^{'}\mu^{'}},\nonumber\\
		&&
	\end{eqnarray}
	where $k_F$ is the Fermi wave vector and the spin operators of itinerant atom $\sigma_{\mu\mu^{'}}$ are
	\begin{eqnarray}
		&&\sigma^z=\frac{1}{2}[3(c^{\dagger}_{\frac{3}{2}}c_{\frac{3}{2}}-c^{\dagger}_{-\frac{3}{2}}c_{-\frac{3}{2}})+(c^{\dagger}_{\frac{1}{2}}c_{\frac{1}{2}}-c^{\dagger}_{-\frac{1}{2}}c_{-\frac{1}{2}})], \nonumber\\
		&&\sigma^{+}=\sqrt{3}(c^{\dagger}_{\frac{3}{2}}c_{\frac{1}{2}}-c^{\dagger}_{-\frac{1}{2}}c_{-\frac{3}{2}})+2c^{\dagger}_{\frac{1}{2}}c_{-\frac{1}{2}}, \nonumber\\
		&&\sigma^{-}=\sqrt{3}(c^{\dagger}_{\frac{1}{2}}c_{\frac{3}{2}}-c^{\dagger}_{-\frac{3}{2}}c_{-\frac{1}{2}})+2c^{\dagger}_{-\frac{1}{2}}c_{\frac{1}{2}}.
	\end{eqnarray}
	Then we get that
	\begin{eqnarray}
		\lefteqn{\hat{H}=\sum\limits_{k,\mu}\epsilon_{k}c_{k,\mu}^{\dagger}c_{k,\mu}}\nonumber\\
		&&-\left(\frac{J}{N}\right)\sum\limits_{k,k^{'}>k_F}
		\{[3(c^{\dagger}_{k\frac{3}{2}}c_{k^{'}\frac{3}{2}}-c^{\dagger}_{k-\frac{3}{2}}c_{k^{'}-\frac{3}{2}}) \nonumber\\
		&&+(c^{\dagger}_{k\frac{1}{2}}c_{k^{'}\frac{1}{2}}-c^{\dagger}_{k-\frac{1}{2}}c_{k^{'}-\frac{1}{2}})]\hat{S}^{z} \nonumber\\
		&&+[\sqrt{3}(c^{\dagger}_{k\frac{1}{2}}c_{k^{'}\frac{3}{2}}
		+c^{\dagger}_{k-\frac{3}{2}}c_{k^{'}-\frac{1}{2}}
		+2c^{\dagger}_{k-\frac{1}{2}}c_{k^{'}\frac{1}{2}})]\hat{S}^{+}\nonumber\\
		&&+[\sqrt{3}(c^{\dagger}_{k\frac{3}{2}}c_{k^{'}\frac{1}{2}}
		+c^{\dagger}_{k-\frac{1}{2}}c_{k^{'}-\frac{3}{2}}
		+2c^{\dagger}_{k\frac{1}{2}}c_{k^{'}-\frac{1}{2}})]\hat{S}^{-}\}. \nonumber\\       	
	\end{eqnarray}
	Under the coupled representation, the common eigenstates $|S,m_{s}\rangle$ can be formed by the spin state $|m_{1},m_{2}\rangle$ of two particles with spin 3/2. They include spin singlet state $\chi_{0,0}$, triplet state $\chi_{1,0 \pm1 }$, quintuplet state $\chi_{2,0 \pm1 \pm2}$ and septuplet state $\chi_{3,0 \pm1 \pm2 \pm3}$,
	\begin{eqnarray}
		&& \chi_{0,0}=\frac{1}{2}\left(	|\frac{3}{2} -\frac{3}{2}\rangle -|\frac{1}{2} -\frac{1}{2}\rangle +|-\frac{1}{2} \frac{1}{2}\rangle -|-\frac{3}{2} \frac{3}{2}\rangle \right), \nonumber\\
		&& \chi_{1,0}=\frac{1}{\sqrt{20}}\left(	3|\frac{3}{2} -\frac{3}{2}\rangle -|\frac{1}{2} -\frac{1}{2}\rangle -|-\frac{1}{2} \frac{1}{2}\rangle +3|-\frac{3}{2} \frac{3}{2}\rangle \right),\nonumber\\
		&& \chi_{1,1}=\frac{1}{\sqrt{10}}\left(	\sqrt{3}|\frac{3}{2} -\frac{1}{2}\rangle -2|\frac{1}{2} \frac{1}{2}\rangle +\sqrt{3}|-\frac{1}{2} \frac{3}{2}\rangle \right),\nonumber\\
		&& \chi_{1,-1}=\frac{1}{\sqrt{10}}\left(\sqrt{3}|-\frac{3}{2} \frac{1}{2}\rangle -2|-\frac{1}{2} -\frac{1}{2}\rangle +\sqrt{3}|\frac{1}{2} -\frac{3}{2}\rangle \right),\nonumber\\
		&& \chi_{2,0}=\frac{1}{2}\left(	|\frac{3}{2} -\frac{3}{2}\rangle +|\frac{1}{2} -\frac{1}{2}\rangle -|-\frac{1}{2} \frac{1}{2}\rangle -|-\frac{3}{2} \frac{3}{2}\rangle \right),\nonumber\\
		&& \chi_{2,2}=\frac{1}{\sqrt{2}}\left(|\frac{3}{2} \frac{1}{2}\rangle -|\frac{1}{2} \frac{3}{2}\rangle \right),\nonumber\\
		&& \chi_{2,-2}=\frac{1}{\sqrt{2}}\left(-|\frac{3}{2} -\frac{1}{2}\rangle -|-\frac{1}{2} -\frac{3}{2}\rangle \right),\nonumber\\
		&& \chi_{2,1}=\frac{1}{\sqrt{2}}\left(|\frac{3}{2} -\frac{1}{2}\rangle -|-\frac{1}{2} \frac{3}{2}\rangle \right),\nonumber\\
		&& \chi_{2,-1}=\frac{1}{\sqrt{2}}\left(-|\frac{3}{2} \frac{1}{2}\rangle -|\frac{1}{2} -\frac{3}{2}\rangle \right),\nonumber\\
		&& \chi_{3,0}=\frac{1}{\sqrt{20}}\left(	|\frac{3}{2} -\frac{3}{2}\rangle +3|\frac{1}{2} -\frac{1}{2}\rangle +3|-\frac{1}{2} \frac{1}{2}\rangle +|-\frac{3}{2} \frac{3}{2}\rangle \right),\nonumber\\
		&& \chi_{3,1}=\frac{1}{\sqrt{5}}\left(|\frac{3}{2} -\frac{1}{2}\rangle + \sqrt{3}|\frac{1}{2} \frac{1}{2}\rangle +|-\frac{1}{2} \frac{3}{2}\rangle \right),\nonumber\\
		&& \chi_{3,-1}=\frac{1}{\sqrt{5}}\left(|-\frac{3}{2} \frac{1}{2}\rangle + \sqrt{3}|-\frac{1}{2} -\frac{1}{2}\rangle +|\frac{1}{2} -\frac{3}{2}\rangle \right),\nonumber\\
		&& \chi_{3,2}=\frac{1}{\sqrt{2}}\left(|\frac{3}{2} \frac{1}{2}\rangle +|\frac{1}{2} \frac{3}{2}\rangle \right),\nonumber\\
		&& \chi_{3,-2}=\frac{1}{\sqrt{2}}\left(|-\frac{3}{2} -\frac{1}{2}\rangle +|-\frac{1}{2} -\frac{3}{2}\rangle \right),\nonumber\\
		&& \chi_{3,3}=|\frac{3}{2} \frac{3}{2}\rangle,
		\chi_{3,-3}=|-\frac{3}{2} -\frac{3}{2}\rangle\nonumber\\
		&&
	\end{eqnarray}
	Due to the energy degeneracy of each configuration, we only need to calculate the bound states with $m_{s}=0$, and the system will choose the one with the lowest energy.	
	A trial wave function $|\psi\rangle$ of the spin-3/2 system consisting of a localized atom and an itinerant atom is assumed to be a combination of spin states. It can be written in the possible ways:
	\begin{eqnarray}
		&&|\psi_7\rangle=\sum_{k>k_F}\Gamma_{k}
		(c^{\dagger}_{k,-\frac{3}{2}}|\frac{3}{2}\rangle
		+3c^{\dagger}_{k,-\frac{1}{2}}|\frac{1}{2}\rangle \nonumber\\
		&&+3c^{\dagger}_{k,\frac{1}{2}}|-\frac{1}{2}\rangle
		+c^{\dagger}_{k,\frac{3}{2}}|-\frac{3}{2}\rangle)|F\rangle,\\
		&&|\psi_5\rangle=\sum_{k>k_F}\Gamma_{k}
		(c^{\dagger}_{k,-\frac{3}{2}}|\frac{3}{2}\rangle
		+c^{\dagger}_{k,-\frac{1}{2}}|\frac{1}{2}\rangle \nonumber\\
		&&-c^{\dagger}_{k,\frac{1}{2}}|-\frac{1}{2}\rangle
		-c^{\dagger}_{k,\frac{3}{2}}|-\frac{3}{2}\rangle)|F\rangle,\\
		&&|\psi_3\rangle=\sum_{k>k_F}\Gamma_{k}
		(3c^{\dagger}_{k,-\frac{3}{2}}|\frac{3}{2}\rangle
		- c^{\dagger}_{k,-\frac{1}{2}}|\frac{1}{2}\rangle \nonumber\\
		&&- c^{\dagger}_{k,\frac{1}{2}}|-\frac{1}{2}\rangle
		+3c^{\dagger}_{k,\frac{3}{2}}|-\frac{3}{2}\rangle)|F\rangle,\\
		&&|\psi_1\rangle=\sum_{k>k_F}\Gamma_{k}
		(c^{\dagger}_{k,-\frac{3}{2}}|\frac{3}{2}\rangle
		-c^{\dagger}_{k,-\frac{1}{2}}|\frac{1}{2}\rangle \nonumber\\
		&&+c^{\dagger}_{k,\frac{1}{2}}|-\frac{1}{2}\rangle
		-c^{\dagger}_{k,\frac{3}{2}}|-\frac{3}{2}\rangle)|F\rangle.
	\end{eqnarray}
	The subscripts 7, 5, 3, and 1 in $|\psi\rangle$ correspond to spin septuplet, quintuplet, triplet, and singlet state, respectively. $\Gamma_{k}$ is a non-zero parameter that needs to be determined. The state $|F\rangle$ designates the Fermi sea of noninteracting particles and
	\begin{eqnarray}\label{F}
		c^{\dagger}_{k\mu}|F\rangle=0, c_{k,mu}|F\rangle=0,	
	\end{eqnarray}
	when T=0K.
	$|\frac{3}{2}\rangle,|-\frac{3}{2}\rangle,|\frac{1}{2}\rangle,|-\frac{1}{2}\rangle,$ represent the localized impurity atom spin states and they satisfy that
	\begin{eqnarray}\label{S}
		&&S^{+}|\frac{3}{2}\rangle=0,
		S^{+}|\frac{1}{2}\rangle=\sqrt{3}|\frac{3}{2}\rangle,\nonumber\\
		&&S^{+}|-\frac{1}{2}\rangle=2|\frac{1}{2}\rangle,
		S^{+}|-\frac{3}{2}\rangle=\sqrt{3}|-\frac{1}{2}\rangle,\nonumber\\
		&&S^{-}|\frac{3}{2}\rangle=\sqrt{3}|\frac{1}{2}\rangle,
		S^{-}|\frac{1}{2}\rangle=2|-\frac{1}{2}\rangle,\nonumber\\
		&&S^{-}|-\frac{1}{2}\rangle=\sqrt{3}|-\frac{3}{2}\rangle,
		S^{-}|-\frac{3}{2}\rangle=0,\nonumber\\
		&&S^{z}|\pm \frac{3}{2}\rangle=\pm \frac{3}{2}|\pm \frac{3}{2}\rangle,
		S^{z}|\pm \frac{1}{2}\rangle=\pm \frac{1}{2}|\pm \frac{1}{2}\rangle,
	\end{eqnarray}
	The eigenvalue equation for the collective spin states formed by the two atoms above the Fermi sea is
	\begin{eqnarray}
		\langle \psi_n|(\hat{H}-E_{n})|\psi_n\rangle=0.
	\end{eqnarray}
	According to Eq.(\ref{F}) and Eq.(\ref{S}), we get the energy equation for the singlet state
	\begin{eqnarray}
		1+30 \left(\frac{J}{N}\right)\sum_{k>k_F}\frac{1}{4(\epsilon_{k}-E_{1})}=0.	
	\end{eqnarray}
	This expression is summed over $k$ and then changed to an integral over the particle energy. The density of states approximately equals to $\rho_{F}$. Finally, the integral gives
	\begin{eqnarray}
		1+\frac{15}{2} J\rho_{F}ln\frac{E_{1}-D}{E_1}=0.
	\end{eqnarray}
	When $|J|\rho_{F}\ll1$, we can get the corresponding energies of these states under antiferromagnetic $(J<0)$ and ferromagnetic coupling $(J>0)$. The solutions to septuplet, quintuplet and triplet case follow the same steps shown in Table 1.
	
\begin{table*}{
		\hbox{\bf  Table 1 The bound state energy of spin-3/2 Fermi gas}}
	{
		\small
		\tabcolsep0.1in
		\begin{tabular}{|c|c|c|c|c|}
			\hline	\hspace{0.2cm}
			{bound state}& singlet & triplet & quintuplet & septuplet  \\
			\hline
			{$J>0$}   & $\frac{D}{1-\exp\left[-\frac{2}{15|J|\rho_F}\right]}$& $\frac{D}{1-\exp\left[-\frac{2}{11|J|\rho_F}\right]}$
			& $\frac{D}{1-\exp\left[-\frac{2}{3|J|\rho_F}\right]}$ & $-D \exp\left[-\frac{2}{9|J|\rho_F}\right]$ \\
			\hline
			{$J<0$}   & $-D \exp\left[-\frac{2}{15|J|\rho_F}\right]$  & $-D \exp\left[-\frac{2}{11|J|\rho_F}\right]$
			& $-D \exp\left[-\frac{2}{3|J|\rho_F}\right]$          &  $\frac{D}{1-\exp\left[-\frac{2}{9|J|\rho_F}\right]}$\\
			\hline 	
		\end{tabular}
	}	
\end{table*}
	It is easily shown from the table that the ferromagnetic s-d coupling $(J>0)$ gives a septuplet bound state whose energy is lower than the Fermi level. If the coupling is antiferromagnetic $(J<0)$, then the energy of the singlet bound state is the lowest. That is to say, for the ferromagnetic coupling, the septuplet state is the most stable state of large spin fermions with spin-3/2. In the case of antiferromagnetic coupling, the ground state is the Kondo singlet state, where the local moments are quenched by the spins of the surrounding itinerant atoms.
	
	Compared with the spin-1/2 system, the ground state energy of the spin-3/2 Fermi gas is lower when the antiferromagnetic coupling constant is the same. It indicates that the larger spin, the easier it was to enter the Kondo screening phase.
	
\section{Conclusion}
	
	In this article, we use the perturbation theory to obtain the scattering probability between a local impurity atom and itinerant atoms in the spin-3/2 Fermi system based on s-d exchange model. We ignored the interaction between the local spin atoms. The resistance anomaly occurs in the system for antiferromagnetic coupling, with the divergence behavior at zero temperature, similar to the results of the spin-1/2 system. However, the spin-3/2 Fermi gas has a higher scattering probability than spin-1/2.
	
	For the study of the ground state of the system, we use a simplified Hamiltonian and only consider the ground state where the z component of total spin is zero. The ground state is also more abundant due to the large-spin Fermi system containing more spin components, which have singlet, triplet, quintuplet, and septuplet states. For the ferromagnetic coupling, except for the septuplet state, the remaining spin-state energies are higher than the Fermi energy, so the septuplet state is the most stable. In the case of antiferromagnetic coupling, the energy of the singlet state is the lowest, so the ground state of the large-spin Fermi system is still Kondo singlet at zero temperature.

\begin{acknowledgements}
Thanks Zhongze Guo and Shuyi Li for their useful suggestions. This study was supported by the National Natural Science Foundation of China (Grant No. 11574028).
\end{acknowledgements}

\end{document}